\documentclass[english,prl,twocolumn,notitlepage,superscriptaddress] {revtex4-1}
\pdfoutput=1
\usepackage[utf8]{inputenc}
\usepackage{graphicx}
\usepackage{amssymb}
\usepackage{setspace}
\usepackage{epstopdf}
\usepackage{multirow}
\usepackage{bm}
\usepackage{braket}
\usepackage{amsmath}
\usepackage{hyperref}

\usepackage{natbib}
\usepackage{graphicx}

\begin{document}
\title{Gate-reflectometry dispersive readout and coherent control of a spin qubit in silicon}

\author{A. Crippa}
\email{alessandro.crippa@cea.fr}
\affiliation{Univ. Grenoble Alpes, CEA, INAC-PHELIQS, F-38000 Grenoble, France}
\author{R. Ezzouch}
\affiliation{Univ. Grenoble Alpes, CEA, INAC-PHELIQS, F-38000 Grenoble, France}
\author{A. Apr\'a}
\affiliation{Univ. Grenoble Alpes, CEA, INAC-PHELIQS, F-38000 Grenoble, France}
\author{A. Amisse}
\affiliation{Univ. Grenoble Alpes, CEA, INAC-PHELIQS, F-38000 Grenoble, France}
\author{R. Lavi\'eville}
\affiliation{CEA, LETI, Minatec Campus, F-38000 Grenoble, France}
\author{L. Hutin}
\affiliation{CEA, LETI, Minatec Campus, F-38000 Grenoble, France}
\author{B. Bertrand}
\affiliation{CEA, LETI, Minatec Campus, F-38000 Grenoble, France}
\author{M. Vinet}
\affiliation{CEA, LETI, Minatec Campus, F-38000 Grenoble, France}
\author{M. Urdampilleta}
\affiliation{Univ. Grenoble Alpes, CNRS, Grenoble INP, Institut N\'eel, F-38000 Grenoble, France}
\author{T. Meunier}
\affiliation{Univ. Grenoble Alpes, CNRS, Grenoble INP, Institut N\'eel, F-38000 Grenoble, France}
\author{M. Sanquer}
\affiliation{Univ. Grenoble Alpes, CEA, INAC-PHELIQS, F-38000 Grenoble, France}
\author{X. Jehl}
\affiliation{Univ. Grenoble Alpes, CEA, INAC-PHELIQS, F-38000 Grenoble, France}
\author{R. Maurand}
\email{romain.maurand@cea.fr}
\affiliation{Univ. Grenoble Alpes, CEA, INAC-PHELIQS, F-38000 Grenoble, France}
\author{S. De Franceschi}
\affiliation{Univ. Grenoble Alpes, CEA, INAC-PHELIQS, F-38000 Grenoble, France}

%\begin{abstract}
%Silicon spin qubits have emerged as a promising path to large-scale quantum processors. In this prospect, the development of scalable qubit readout schemes involving a minimal device overhead is a compelling step. Here we report the implementation of gate-coupled rf reflectometry for the dispersive readout of a fully functional spin qubit device. We use a p-type double-gate transistor made using industry-standard silicon technology. The first gate confines a hole quantum dot encoding the spin qubit, the second one a helper dot enabling readout. The qubit state is measured through the phase response of a lumped-element resonator to spin-selective interdot tunneling. The demonstrated qubit readout scheme requires no coupling to a Fermi reservoir, thereby offering a compact and potentially scalable solution whose operation may be extended above 1 Kelvin.
%\end{abstract}

\maketitle

\textbf{Silicon spin qubits have emerged as a promising path to large-scale quantum processors. In this prospect, the development of scalable qubit readout schemes involving a minimal device overhead is a compelling step. Here we report the implementation of gate-coupled rf reflectometry for the dispersive readout of a fully functional spin qubit device. We use a p-type double-gate transistor made using industry-standard silicon technology. The first gate confines a hole quantum dot encoding the spin qubit, the second one a helper dot enabling readout. The qubit state is measured through the phase response of a lumped-element resonator to spin-selective interdot tunneling. The demonstrated qubit readout scheme requires no coupling to a Fermi reservoir, thereby offering a compact and potentially scalable solution whose operation may be extended above 1\,K.}\\

The recent years have witnessed remarkable progress in the development of semiconductor spin qubits \cite{petta2005coherent, LPK_CNTqubit, diamond, HannesGe} with an increasing focus on silicon-based realizations \cite{kawakami, roro, HRLsymmetricOperation, Hybrid_extending}.
Access to isotopically enriched $^{28}\text{Si}$ has enabled the achievement of very long spin coherence times for both nuclear and electron spins \cite{VeldhorstAddressable, Morello30seconds, Yoneda99}. In addition, two-qubit gates with increasing high fidelities were demonstrated in electrostatically defined electron double quantum dots \cite{PettaCNOT, Watson, Dzurak-2qubitFidelity}.\\
While further improvements in single- and two-qubit gates can be expected, growing research efforts are now being directed to the realization of scalable arrays of coupled qubits \cite{MennoCMOSarchitecture, RoyArchitecture, JonesArchitecture, ZajacArray, hotdensecoherent}.
Leveraging the well-established silicon technology may enable facing the scalability challenge, and initiatives to explore this opportunity are on the way \cite{LouisESSDERC2018}. Simultaneously, suitable qubit device geometries need to be developed. One of the compelling problems is to engineer scalable readout schemes. The present work addresses this important issue.\\  
It has been shown that a microwave excitation applied to a gate electrode drives Rabi oscillations via the electric-dipole spin resonance mechanism \cite{kawakami, HannesGe, Nowack07, PeterssonQED, roro, gtensor}. The possibility of using a gate as sensor for qubit readout would allow for a compact device layout, with a clear advantage for scalability. Gate reflectometry probes charge tunneling transitions in a quantum dot system through the dispersive shift of a radiofrequency (rf) resonator connected to a gate electrode \cite{Ciccarelli, Colless2013dispersive, Betz_limit, CrippaNL}. Jointly to spin-selective tunneling, e.g. due to Pauli spin Blockade in a double quantum dot (DQD), this technique provides a way to measure spin states.\\
In a similar fashion, the phase shift of a superconducting microwave resonator coupled to the source of an InAs nanowire has enabled spin qubit dispersive readout \cite{PeterssonQED}.\\
In Si, recent gate reflectometry experiments have shown single-shot electron spin detection \cite{SimmonsSingleshot, DzurakSingleshot, MatiasSingleshot}. 
Here, we combine coherent spin control and gate dispersive readout in a compact qubit device.
%Recent experiments have shown single-shot electron spin readout in Si by gate reflectometry \cite{SimmonsSingleshot, DzurakSingleshot, MatiasSingleshot}. Here, we implement rf gate reflectometry in a fully functional spin qubit device with a compact layout. 
Two gates tune an isolated hole DQD, and two distinct electric rf tones (one per gate) allow spin manipulation and dispersive readout.
Spin initialization and control are performed without involving any charge reservoir; qubit readout relies on the spin-dependent phase response at the DQD charge degeneracy point.
We assess hole single spin dynamics and show coherent spin control, validating a protocol for complete qubit characterization exploitable in more complex architectures.\\
%\section{Figure1}
\begin{figure}
\centering
	\includegraphics[width=\columnwidth]{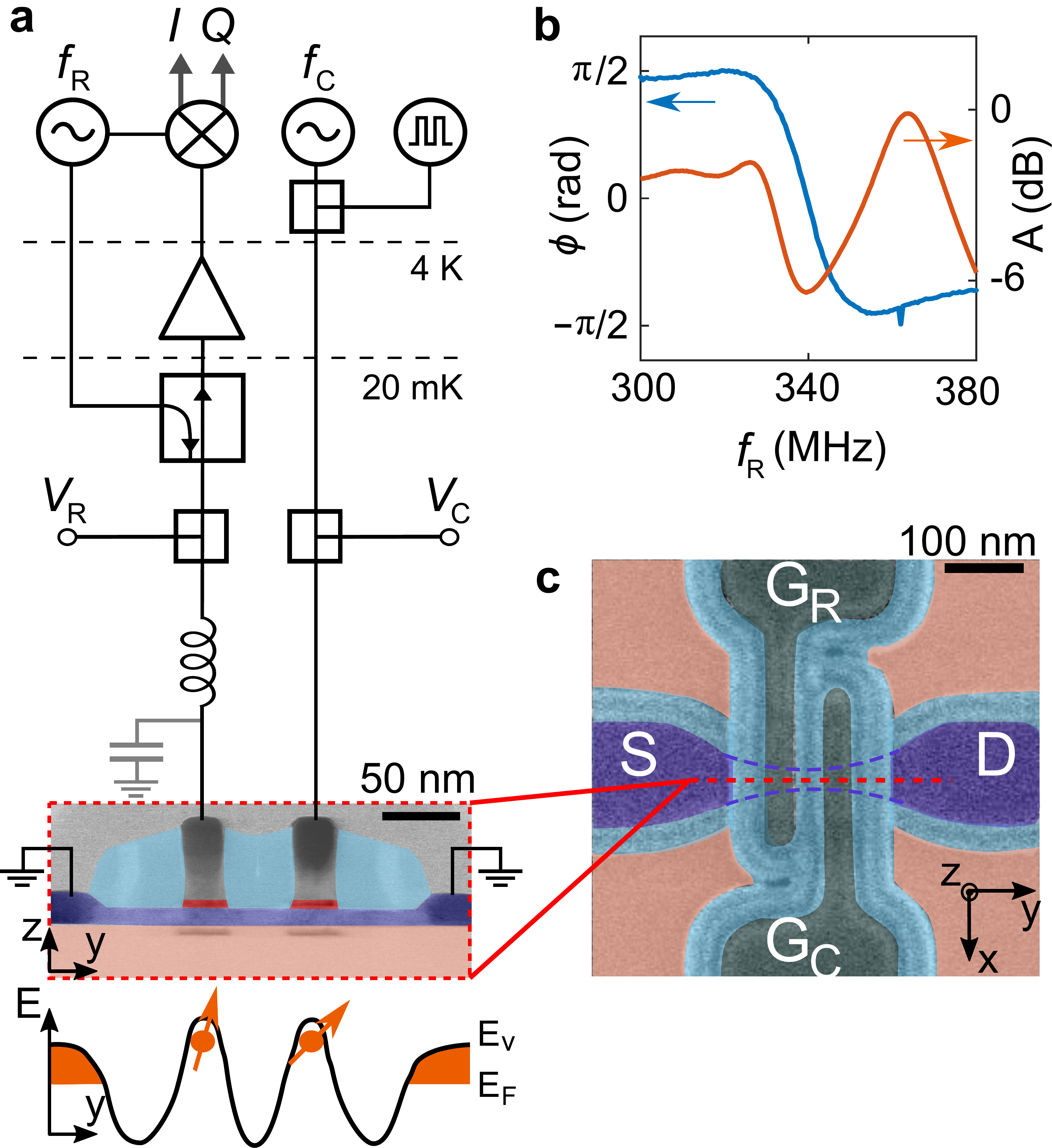} 
	\caption{\textbf{Device layout and circuitry for qubit dispersive sensing and manipulation.} (a) False-color transmission electron micrograph of a double-gate Si device. The 11-nm-thick, 35-nm-wide Si nanowire (light blue) connects p-type, boron-doped source-drain contacts (dark blue) and lies on a 140-nm-thick $\text{SiO}_2$ buffer layer (pink). The two 35-\,nm-wide gates (gray) are spaced by 35\,nm. $\text{Si}_3 \text{N}_4$ spacers (cyan) prevent dopant implantation in the Si channel. At 20\,mK, proper gate voltages induce the accumulation of two hole quantum dots: one can be used as a spin qubit, the other as a helper dot for qubit readout. One gate is connected to a lumped-element resonator excited at frequency $f_R$ for dispersive readout. 
    A ultra-high frequency digital lock-in demodulates the reflected signal after a directional coupler, separating the incoming and outgoing waves, and a low-noise amplifier at 4\,K. 
    The other gate applies square pulses and GHz radiation to drive controlled coherent rotations of the hole spin qubit. At the bottom, DQD energy diagram with $E_v$ as valence band edge and $E_F$ as Fermi energy.
    (b) Phase response ($\phi$) and attenuation ($A$) of the resonator at base temperature. (c) Scanning electron micrograph of the device.}
    	\label{fig:Fig1}
\end{figure}
\\
%\textbf{Results}\\
%\textbf{Double quantum dot dispersive spectroscopy.} 
The experiment is carried out on a double-gate, p-type Si transistor fabricated on a silicon-on-insulator 300-mm wafer using an industry-standard fabrication line \cite{roro}.
The device, nominally identical to the one in Fig.\,\ref{fig:Fig1}c, has two parallel top gates, $\text{G}_{\text{R}}$ and $\text{G}_{\text{C}}$, wrapping an etched Si nanowire channel. The gates are defined by e-beam lithography and have enlarged overlapping spacers to avoid doping implantation in the channel. The measurement circuit is shown in Fig.\,\ref{fig:Fig1}a.
At low temperature (we operate the device at 20$\,$mK using a dilution refrigerator), DC voltages $V_C$ and $V_R$ are applied to these gates to induce two closely spaced hole quantum dots. The 'Control' gate $\text{G}_{\text{C}}$ delivers also sub-$\mu$s pulses and microwave excitation in the GHz range to manipulate the qubit. 
The 'Readout' gate, $\text{G}_{\text{R}}$, is wire-bonded to a 220$\,$nH surface-mount inductor. Along with a parasitic capacitance and the device impedance, the inductor forms a tank circuit resonating at $f_0 = 339$\,MHz. Figure\,\ref{fig:Fig1}b shows the phase $\phi$ and attenuation $A$ of the reflected signal as a function of the resonator driving frequency $f_R$. From the slope of the phase trace at $f_0$ we extract a quality factor $Q_{\text{loaded}} \simeq 18$. The qubit device acts as a variable impedance load for the resonator, and the resonant frequency $f_0$ undergoes a dispersive shift according to the state of the qubit.
%\section{Figure 2}
\begin{figure*}
\centering
	\includegraphics[width=2\columnwidth]{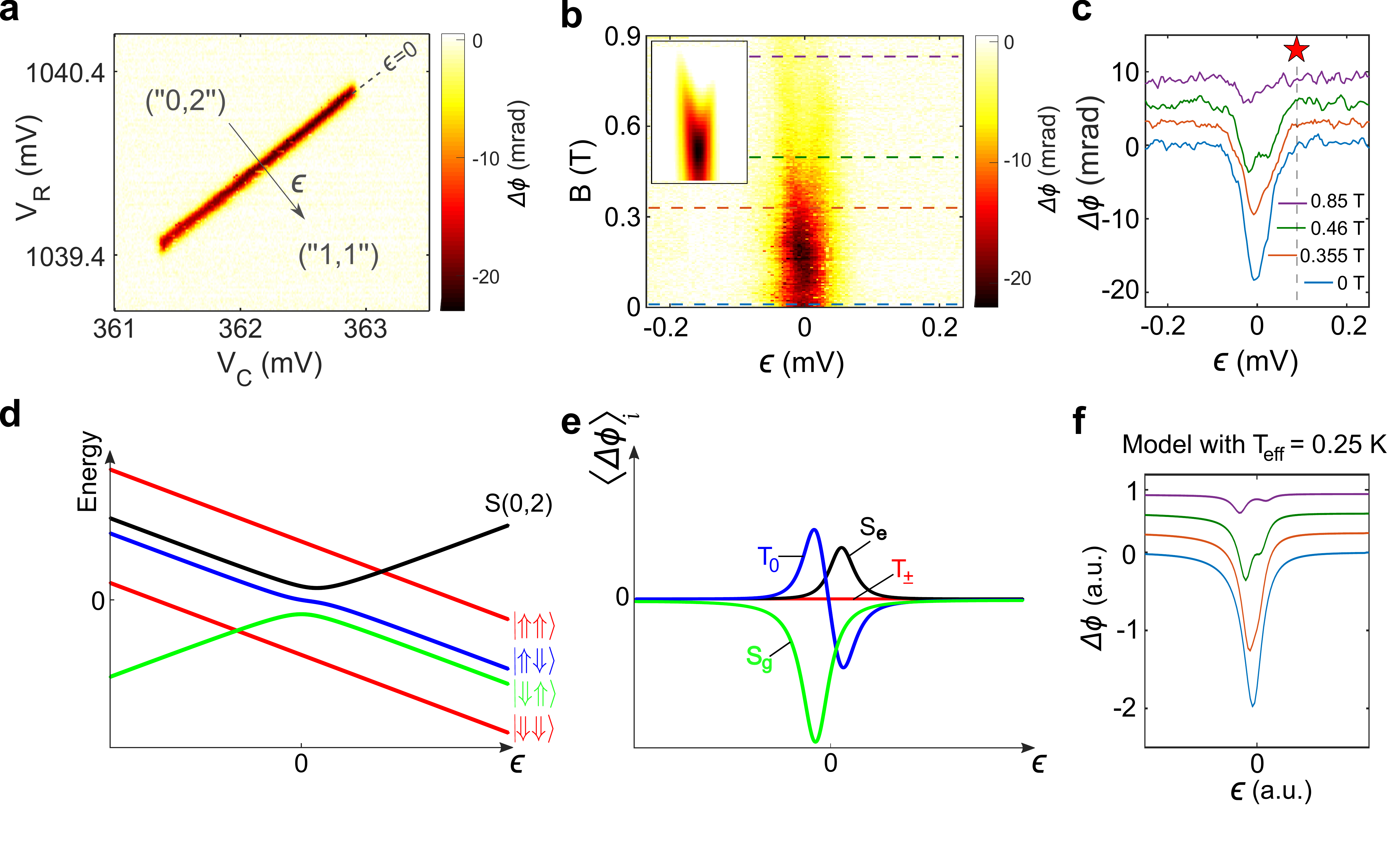} 
	\caption{\textbf{Magnetospectroscopy of the double quantum dot.} (a) Phase shift of the reflected signal as a function of $V_C$ and $V_R$ near the interdot transition line under study. The arrow indicates $\epsilon$ detuning axis. (b) Interdot dispersive signal as a function of a magnetic field $B$ oriented along the nanowire axis. The phase response diminishes with $B$, denoting a interdot charge transition of $(0,2) \leftrightarrow (1,1)$ type. Inset: theoretical prediction of the dispersive response for a DQD model taking into account thermal spin populations, see Supplementary Note 3. (c) Line cuts of the plot in panel b) at the position of the dashed lines. Data are offset for clarity. (d) Schematic of the DQD energy levels close to a $(0,2) \leftrightarrow (1,1)$ transition at finite $B$ and for $|g^*_L-g^*_R|=0.5$  (e) Thermally-averaged phase response $\langle \Delta \phi \rangle _i$ with $T_\text{eff}= 0.25$\,K. $\langle \Delta \phi \rangle_i$ is second derivative of the energy-level dispersion of state $i$ in panel d), weighted by the occupation probability. Here $i$ labels the different DQD states, i.e. the singlets $S_g$ (green) and $S_e$ (black), and the triplets $T_0$ (blue), $T_-$ (red), and $T_+$ (red). (f) Qualitative phase shift resulting from the sum of all $\langle \Delta \phi \rangle_i$ from panel e). A double-peak structure emerges at sufficiently high $B$ in qualitative agreement with the experimental data in panel c).}
	\label{fig:Fig2}
\end{figure*}
\\
To determine the charge stability diagram of our DQD, we probe the phase response of the resonator while sweeping the DC gate voltages $V_R$ and $V_C$ (see Supplementary Note 2 and Supplementary Fig.\,2).\\ 
The diagonal ridge in Fig.\,\ref{fig:Fig2}a denotes the interdot charge transition we shall focus on hereafter. 
Along this ridge, the electrochemical potentials of the two dots line up enabling the shuttling of a hole charge from one dot to the other. This results in a phase variation $\Delta \phi$ in the reflected signal. Quantitatively, $\Delta \phi$ is proportional to the quantum capacitance associated with the gate voltage dependence of the energy levels involved in the interdot charge transition. Interdot tunnel coupling results in the formation of molecular bonding ($+$) and anti-bonding ($-$)  states with energy levels $E_+$ and $E_-$, respectively. These states have opposite quantum capacitance since $C_{Q,\pm}=-\alpha^2 (\partial^2 E_{\pm}/\partial \epsilon^2)$ \cite{CrippaNL}. Here $\epsilon$ is the gate-voltage detuning along a given line crossing the interdot charge transition boundary, and  $\alpha$ is a lever-arm parameter relating $\epsilon$ to the energy difference between the electrochemical potentials of the two dots (we estimate $\alpha \simeq 0.58$\,eV V$^{-1}$ along the detuning line in Fig.\,\ref{fig:Fig2}a). The width of the $\Delta \phi$ ridge, once translated into energy, gives the interdot tunnel coupling, $t$. We estimate $t$ between 6.4 and 8.5\,$\mu$eV, depending on whether thermal fluctuations contribute or not to the dispersive response (see Supplementary Note 3).\\
The total charge parity and the spin character of the DQD states can be determined from the evolution of the interdot ridge in an applied magnetic field, $B$ \cite{Petta_parity}. Figure\,\ref{fig:Fig2}b shows the $B$-dependence of the phase signal at the detuning line indicated in Fig.\,\ref{fig:Fig2}a. Four representative traces taken from this plot are shown in Fig.\,\ref{fig:Fig2}c. 
The interdot phase signal progressively drops with $B$. At $B = 0.355$\,T the line profile is slightly asymmetric, while a double-peak structure emerges at $B = 0.46$\,T. The two peaks move apart and weaken by further increasing $B$, as revealed by the trace at $B=0.85$\,T.\\
The observed behavior can be understood in terms of an interdot charge transition with an even number of holes in the DQD, in a scenario equivalent to a  $(0,2) \leftrightarrow (1,1)$ transition. We shall then refer to a ''(0,2)'' and a ''(1,1)'' state, even if the actual number of confined holes is larger (we estimate around ten, see Supplementary Note 2). 
The $\epsilon$ dependence of the DQD states at finite $B$ is presented in Fig.\,\ref{fig:Fig2}d. 
Deeply in the positive detuning regime, different $g$-factors for the left ($g_L^*$) and the right dot ($g_R^*$) result in four non-degenerate  $(1,1)$ levels corresponding to the following spin states: $\ket{\Downarrow \Downarrow}$, $\ket{\Uparrow \Downarrow}$, $\ket{\Downarrow \Uparrow}$, $\ket{\Uparrow \Uparrow}$ \cite{LPKSOqubit, LPK_PRL12, PeterssonQED}. At large negative detuning, the ground state is a spin-singlet state $S(0,2)$ and the triplet states $T(0,2)$ lie high up in energy. Around zero detuning, the $\ket{\Uparrow \Downarrow}$, $\ket{\Downarrow \Uparrow}$ states hybridize with the $S(0,2)$ state forming an unpolarized triplet $T_0(1,1)$ and two molecular singlets, $S_{g}$ and $S_{e}$, with bonding and anti-bonding character, respectively (Supplementary Note 3).\\
We use the spectrum of Fig.\,\ref{fig:Fig2}d to model the evolution of the interdot phase signal in Fig.\,\ref{fig:Fig2}b-c.
Importantly, we make the assumption that the average occupation probability of the available excited states are populated according to a Boltzmann distribution with an effective temperature $T_{\text{eff}}$, which is used as a free parameter. Because the reflectometry signal is averaged over many resonator cycles, $\Delta \phi = \sum_i \langle \Delta \phi \rangle _i$, where $\langle \Delta \phi \rangle_i$ is the phase response associated to state $i$ weighted by the respective occupation probability \cite{Petta_parity} (here $i$ labels the DQD levels in Fig.\,\ref{fig:Fig2}d). Figure\,\ref{fig:Fig2}e shows $\langle \Delta \phi \rangle _i$ as a function of $\epsilon$ for $T_{\text{eff}}=250$\,mK. The spin polarized triplet states $T_-$ and $T_+$  (i.e. $\ket{\Downarrow \Downarrow}$ and  $\ket{\Uparrow \Uparrow}$, respectively) are linear in $\epsilon$ and, therefore, they do not cause any finite phase shift; $S_g$, $S_e$, and $T_0(1,1)$, on the other hand, possess a curvature and are sensed by the reflectometry apparatus (Supplementary Note 3). We note that the phase signal for $T_0(1,1)$ has a peak-dip line shape whose minimum lies at positive $\epsilon$ (blue trace), partly counterbalanced by the positive phase signal due to $S_e$. The $S_g$ state causes a pronounced dip at negative $\epsilon$ (green trace), dominating over the peak component of $T_0$.  
The overall net result is a phase signal with an asymmetric double-dip structure consistent with our experimental observation.\\
This simple model, with the chosen $T_{\text{eff}}=250$\,mK, qualitatively reproduces the emergence of the double-dip structure at $B \sim 0.4 $\,T, as well as its gradual suppression at higher $B$, as shown in the inset to Fig.\,\ref{fig:Fig2}b and in Fig.\,\ref{fig:Fig2}f (increasing the Zeeman energy results in the depopulation of the $S_g$ and $T_0$ excited states in favor of the $T_-(1,1)$ ground state, for which $\Delta \phi=0$).\\
%ground state following the depopulation of the reflecting the $T_-(1,1)$ becomes the only occu (see.   double-dip  In our case $T_0(1,1)$ and $S_g$ are populated and appear in Fig.\,\ref{fig:Fig2}b as two faint branches discernable from $B \sim 0.4 $\,T. At such a field, having $T_-(1,1)$ occupied with unit probability would imply $\Delta \phi=0$ according to the schematics of Fig.\,\ref{fig:Fig2}d.\\ 
%\RED{\sout{The isolation of the DQD from the reservoirs hampers an absolute estimation of the hole temperature in the DQD. Defining an effective temperature $T_{\text{eff}}$ such that $T_{\text{eff}} \gtrsim t/k_B$ ($k_B$ is the Boltzman constant), with $T_{\text{eff}}=0.25$\,K}}.
%With $T_{\text{eff}}=250$\,mK we obtain a magnetic field dependency of the interdot phase signal in qualitative agreement with the experimental data (inset in Fig.\,\ref{fig:Fig2}b); Figure\,\ref{fig:Fig2}e also captures the asymmetric line shape at low field and the appearance of the phase signal from the states $S_g$ and $T_0$ at $B>0.4$\,T. 
%\RED{\sout{Such a scenario predicts also the twofold effect of increasing the magnetic field in Fig.\,\ref{fig:Fig2}b: first, the suppression of the phase shift occurs as $T_-(1,1)$ gets lowered in energy so that the thermal pumping becomes progressively inefficient; second, the pair of peaks associated to the singlets and $T_0(1,1)$ moves apart in $\epsilon$. In Fig.\,\ref{fig:Fig2}f we plot the phase response predicted for the field values of panel c) considering $T_{\text{eff}}=0.25$\,K; a qualitative agreement supports our model.}}\\
%\section{Figure 3}
\begin{figure}
\centering
	\includegraphics[width=0.82\columnwidth]{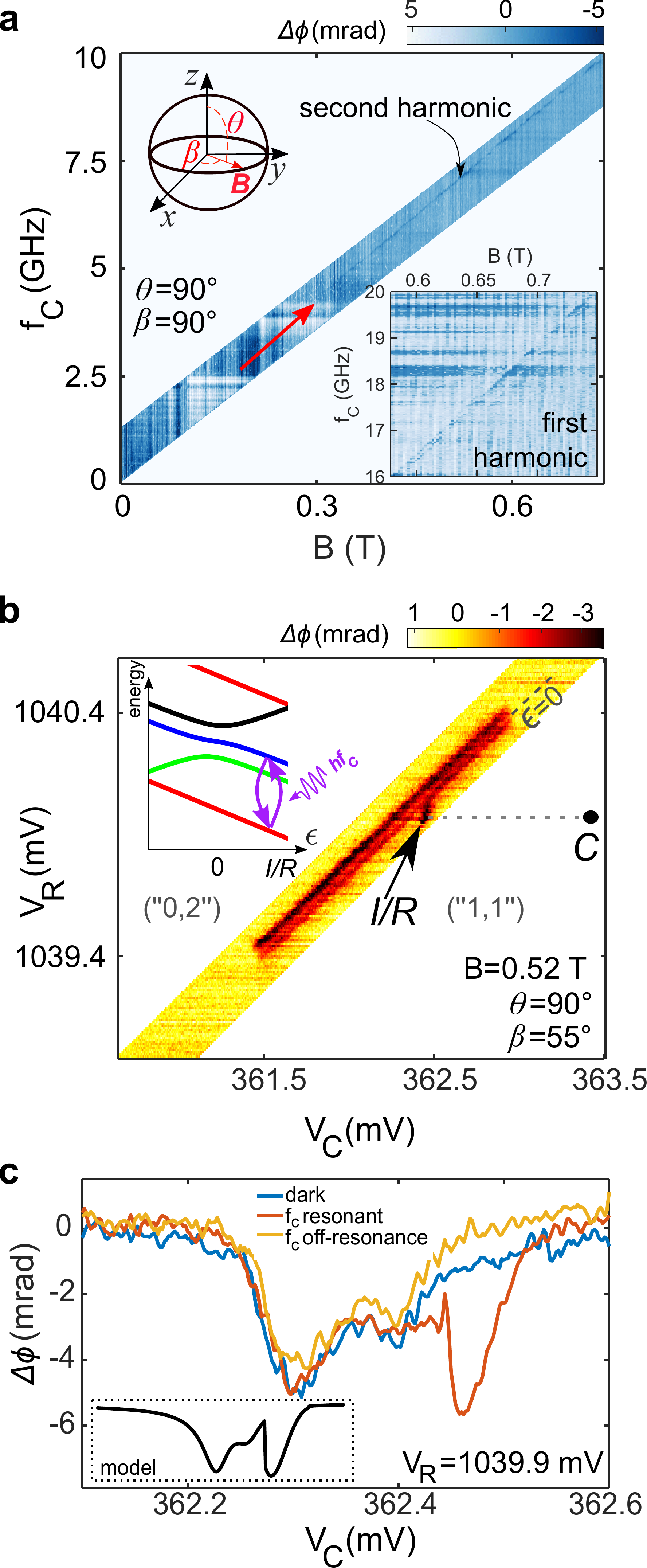} 
	\caption{\textbf{Experimental detection of  electric-dipole spin resonance (EDSR).} (a) Phase response as a function of $B$  and microwave frequency $f_C$. $B$ is oriented along the $y$ direction with respect to the frame of Fig.\,\ref{fig:Fig1}a. The linear phase ridge denoted by a red arrow is a characteristic signature of EDSR. It corresponds to a second-harmonic signal, while the much weaker first harmonic is shown in the lower inset. 
	%The phase of the reflected signal changes when the energy of the photon matches the Zeeman splitting of spin states with different $C_Q$. 
 (b) Stability diagram at $B=0.52$\,T (orientation $\beta=55^{\circ}$ and $\theta=90^{\circ}$ according to the diagram in upper inset of a) with $f_C = 7.42$\,GHz and microwave power $P_C \approx -80$\,dBm. EDSR between $T_-(1,1)$ and $T_0(1,1)$ (purple arrows in inset) is driven at point $I$. In the stability diagram, the change of population induced by EDSR is visible as a localized phase signal at point $I/R$. (c) Phase shift at $V_R = 1039.9$\,mV as a function of $V_C$ without microwave irradiation (dark), and under on-resonance and off-resonance excitation at $f_C=7.42$ and 7.60 \,GHz, respectively. EDSR-stimulated transitions appear as a pronounced peak whose position and line shape are compatible with our model (inset).}
	\label{fig:Fig3}
\end{figure}
\\
%\textbf{Dispersive detection of electric-dipole spin resonance.}
Now that we have elucidated the energy level structure of the DQD, we can discuss the operation of the device as a single-hole spin qubit with electrical control and dispersive readout. %Together with initialization, they represent the fundamental operations to use our transistor as a spin qubit.
Electric dipole spin resonance (EDSR) \cite{roro, gtensor, Benjamin} is induced by a microwave voltage modulation applied to gate $G_C$. To detect EDSR dispersively, the resonating states must have different quantum capacitances. The DQD is initially tuned to the position of the red star in Fig.\,\ref{fig:Fig2}c, where the DQD is in a "shallow" (1,1) configuration, i.e. close to the boundary with the (0,2) charge state (more details in Supplementary Note 4 and Supplementary Fig.\,4).\\
Figure\,\ref{fig:Fig3}a shows the dispersive measurement of an EDSR line. The microwave gate modulation of frequency $f_C$ is applied continuously and $B$ is oriented along the nanowire axis. 
%$hf_C=g\mu_BB$ and the dispersive signal changes as the population of the states is modified.
We ascribe the resonance line to a second harmonic driving process where  2$hf_C=g\mu_BB$ ($h$ the Planck's constant, $\mu_B$ the Bohr magneton and $g$ the effective hole $g$-factor). From this resonance condition we extract $g=1.735 \pm 0.002$, in agreement with previous works \cite{roro, gtensor}. The first harmonic signal, shown in the inset to Fig.\,\ref{fig:Fig3}a, is unexpectedly weaker. Though both first and second harmonic excitations can be expected \cite{ScarlinoSecond}, the first harmonic EDSR line (inset to Fig.\,\ref{fig:Fig3}a) is unexpectedly weaker. A comparison of the two signal intensities requires the knowledge of many parameters (relaxation rate, microwave power, field amplitude) and calls for deeper investigations.\\
The visibility of the EDSR signal can be optimized by a fine tuning of the gate voltages. Figure\,\ref{fig:Fig3}b shows a high-resolution measurement over a narrow region of the stability diagram around the interdot charge transition boundary at $B=0.52$\,T; the interdot line has a double peak structure, consistently with the data in Fig.\,\ref{fig:Fig2}b-c. The measurement is performed while applying a continuous microwave tone $f_C=7.42$\,GHz. EDSR appears as a distinct phase signal around $V_C\simeq 362.5$\,mV and $V_R \simeq 1040$\,mV, i.e. slightly inside the $(1,1)$ charge region, pinpointed by the black arrow as $I/R$. Such EDSR feature is extremely localized in the stability diagram reflecting the gate-voltage dependence of the hole $g$-factor \cite{gtensor}.\\
Figure\,\ref{fig:Fig3}c displays line cuts across the interdot transition line at fixed $V_R$ and different microwave excitation conditions. With no microwaves excitation, we observe the double-peak line shape discussed above. With a microwave gate modulation at $f_C=7.42$\,GHz, the spin resonance condition is met at $V_C\simeq362.45$\,mV, which results in a pronounced EDSR peak, the same observed at point $I/R$ in Fig.\,\ref{fig:Fig3}b (see also Supplementary Fig.\,4). The peak vanishes when $f_C$ is detuned by 20\,MHz (cyan trace).\\
At point $I/R$, resonant microwave excitation enables the spectroscopy of the $T_0(1,1)$ state.
The inset to Fig.\,\ref{fig:Fig3}c shows the signal we expect from our model (Supplementary Note 4). In a small detuning window, the populations of $T_-(1,1)$ and $T_0(1,1)$ are assumed to be balanced by EDSR (see the energy levels in the inset to Fig.\,\ref{fig:Fig3}b); this results in a phase signal dramatically enhanced resembling the feature centered at $I/R$ in the main panel. A further confirmation that the spin transitions are driven between $T_-(1,1)$ and $T_0(1,1)$ is given by the extrapolated intercept at 0\,T of the EDSR transition line in Fig.\,\ref{fig:Fig3}a, found much smaller ($<100$\,MHz) than $t$. In the following, we shall use point $I/R$ to perform qubit initialization and readout.\\
\begin{figure}
\centering
	\includegraphics[width=\columnwidth]{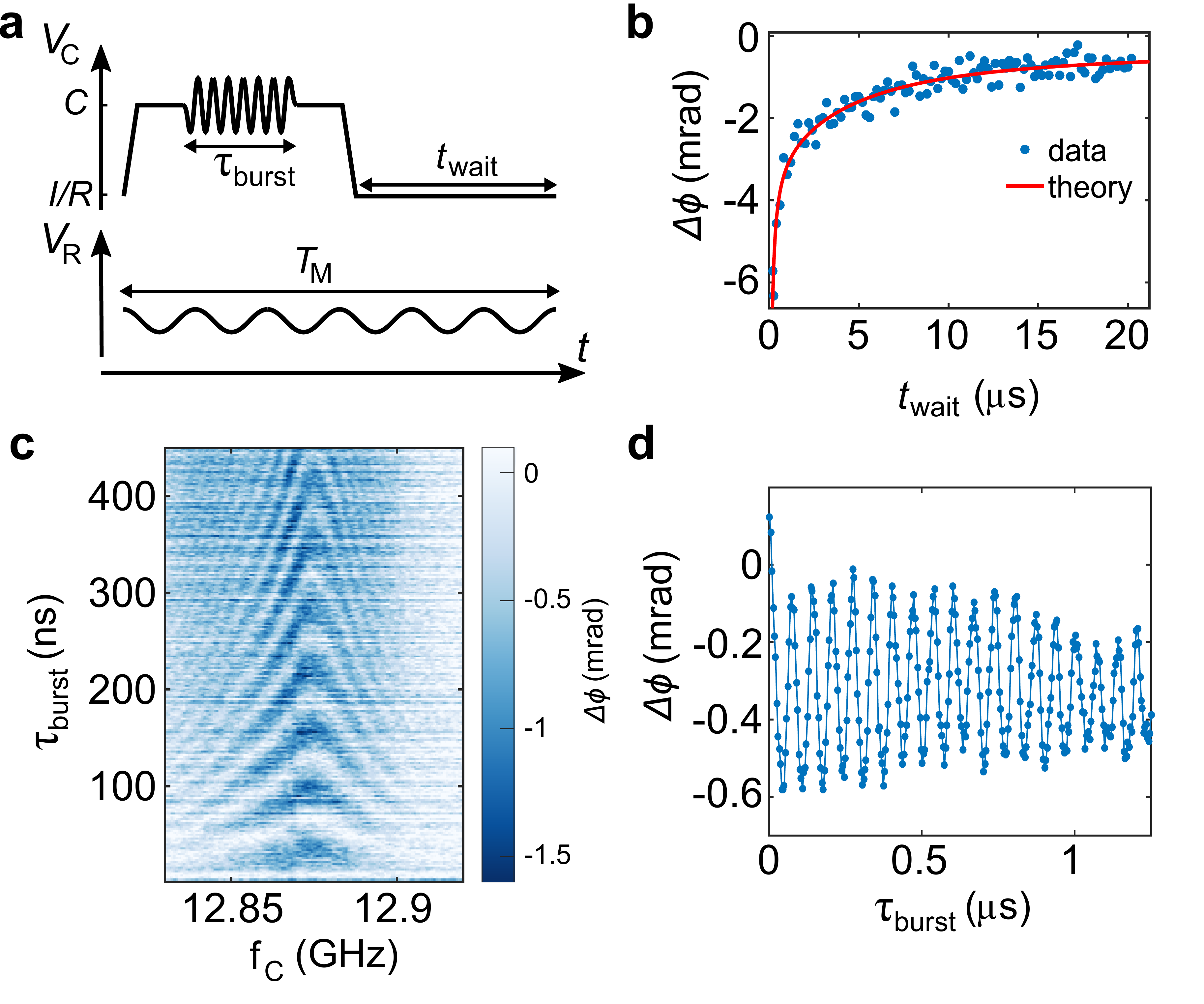} 
	\caption{\textbf{Single spin control and dispersive sensing.} (a) The pulse sequence alternating between "deep" $(1,1)$ regime ($C$) for spin manipulation and "shallow" $(1,1)$ regime ($I/R$), close to the $(0,2) \leftrightarrow (1,1)$ transition, for the readout and resetting of the spin system. A microwave burst rotates the hole spin during the manipulation stage. The readout tone is continuously applied throughout the sequence period $T_M$. (b) Phase shift as a function of $t_{\text{wait}}$ for a $\simeq 1$\,mV pulse on $V_C$ with $\tau_{\text{burst}}=100$\,ns and $f_C=12.865$\,GHz, with $B = 0.512$\,T along $\beta=0^{\circ}$ and $\theta=60^{\circ}$. The phase signal approaches 0 when $t_{\text{wait}} \gg T_1$. A simple model yields $T_1=2.7 \pm 0.7$\,$\mu$s. 
	%(c) Rabi experiment
	(c) Dispersive signal $\Delta \phi\,(f_C, \tau_{\text{burst}})$, measured with the detuning pulses of panel a) with $t_{\text{wait}}= 1$\,$\mu$s. Four maps have been averaged. 
	(d) Phase response as a function of EDSR burst time at $f_C=12.865$\,GHz. The plot shows Rabi oscillations with 15\,MHz frequency due to coherent spin rotations. Each data point is integrated for 100\,ms and then averaged over 30 traces.}
	\label{fig:Fig4}
\end{figure}
\\
%\textbf{Qubit control and readout.}
The device is operated as a spin qubit implementing the protocol outlined in Fig.\,\ref{fig:Fig4}a. The voltage sequence in the upper part of Fig.\,\ref{fig:Fig4}a %is superimposed to $V_C = 362.5$\,mV to prepare, control and initialize a single hole spin. 
tunes the DQD at the control point $C$ ($\simeq 1$\,mV deep in the $(1,1)$ region) where holes are strongly localized in either one or the other dot with negligible tunnel coupling. A microwave burst of duration $\tau_{\text{burst}}$ and frequency $f_C$ drives single spin rotations between $\ket{\Downarrow \Downarrow}$ and $\ket{\Uparrow \Downarrow}$; the system is then brought back to $I/R$ in the "shallow" $(1,1)$ regime for a time $t_{\text{wait}}$ for readout and initialization. The dispersive readout eventually relies on the spin-resolved phase shift at $I/R$, though the reflectometry tone $f_R$ is applied during the whole sequence period $T_M$ and the reflected signal is streamed constantly to the acquisition module.\\
First, we determine the lifetime $T_1$ of the excited spin state at the readout point $I/R$ by sweeping $t_{\text{wait}}$ after a $\pi$-burst at point $C$. 
The results are shown in Fig.\,\ref{fig:Fig4}b. The phase signal rapidly diminishes with increasing $t_{\text{wait}}$ because spin relaxation depopulates the excited spin state in favor of the non-dispersive $T_-(1,1)$ ground state. The estimated spin lifetime at the readout position is $T_1=2.7 \pm 0.7$\,$\mu$s (see Supplementary Note 5). By shifting the position of a 100\,ns microwave burst within a 12\,$\mu$s pulse, no clear decay of the dispersive signal is observed, which suggests a spin lifetime at manipulation point longer than 10\,$\mu$s.\\
We demonstrate coherent single spin control in the chevron plot of Fig.\,\ref{fig:Fig4}c. The phase signal is collected as a function of microwave burst time $\tau_{\text{burst}}$ and driving frequency $f_C$.
The spin state is initialized at point $I/R$ ($t_{\text{wait}} \sim T_1$). In Fig.\,\ref{fig:Fig4}d the phase signal is plotted as a function of $\tau_{\text{burst}}$ with $f_C$ set at the Larmor frequency $f_{\text{Larmor}}$. The Rabi oscillations have 15\,MHz frequency, consistent with Refs.\,\citenum{roro} and \citenum{gtensor}.
The non-monotonous envelope is attributed to random phase accumulation in the qubit state by off-resonant driving at $f_{\text{Larmor}} \pm f_R$ due to up-conversion of microwave and reflectometry tones during the manipulation time. 
Data in Fig.\,\ref{fig:Fig4}d have been averaged over 30 measurements though the oscillations are easily distinguishable from single scans where each point is integrated over 100\,ms. 
Figure\,\ref{fig:Fig4} witnesses the success of using electrical rf signals both for coherent manipulation by EDSR and for qubit-state readout by means of gate reflectometry.\\
\\
%\textbf{Discussion}\\
The measured $T_1$ is compatible with the relaxation times obtained for hole singlet-triplet states in acceptor pairs in Si \cite{RoggeT1} and in Ge/Si nanowire double quantum dots \cite{NLholesFerdinand}; in both cases $T_1$ has been measured at the charge degeneracy point with reflectometry setups similar to ours.
Nonetheless, charge detector measurements have shown $T_1$ approaching 100\,$\mu$s for single hole spins in Ge hut wire quantum dots \cite{LadaSingleShot} and $\lesssim 1$\,ms for Ge/Si singlet-triplet systems \cite{NatNanoHoleFerdinand}. This suggests that despite the intrinsic spin-orbit coupling single spin lifetimes in the ms range might be achievable in Si too.
Strategies to boost $T_1$ at the readout point may consist of inserting rf isolators between the coupler and the amplifier to reduce the backaction on the qubit and avoiding high-$\kappa$ dielectric in the gate stack to limit charge noise.\\
We note that $T_1$ could depend on the orientation of the magnetic field as well \cite{SpinOrbit_donors}. Future studies on magnetic field anisotropy will clarify whether $T_1$, along with the effective $g$-factors (and hence the dispersive shift for readout) and Rabi frenquency, can be maximized at once along a specific direction.
Technical improvements intended to enhance the phase sensitivity, like resonators with higher Q-factor and parametric amplification, could push the implemented readout protocol to distinguish spin states with a micro-second integration time, enabling single shot measurement as reported in a recent experiment with a gate-connected superconducting resonant circuit \cite{Lieven2019rapid}. Lastly, the resonator integration in the back-end of the industrial chip could offer the possibility to engineer the resonant network at a wafer scale, guaranteeing  controlled and reproducible qubit-resonator coupling.\\
The gate-based dispersive sensing demonstrated here does not involve local reservoirs of charges or embedded charge detectors. This meets the requirements of forefront qubit architectures (e.g. Ref.\,\citenum{RoyArchitecture}), where the spin readout would be performed at will by any gate of the 2D quantum dot array by frequency multiplexing.\\
Dispersive spin detection by Pauli blockade has a fidelity not constrained by the temperature of the leads. As recently shown \cite{Dzurak1Kelvin}, isolated DQDs can serve as spin qubits even if placed at environmental temperatures exceeding the spin splitting, like 1\,K or more. This should relax many cryogenic constraints and support the co-integration with classical electronics, as required by a scale-up perspective \cite{hotdensecoherent}.

\section{Methods}
\textbf{Device fabrication.} The fabrication process of the device was carried out in a 300\,mm CMOS platform and is described in Ref.\,\citenum{roro}.\\
\textbf{Experimental set-up.} The experiment is performed by exciting the resonator input at $f_R = f_0 = 339$\,MHz and power $P_R \approx -110$\,dBm. We measure the phase variation $\Delta \phi$ of the reflected signal isolated from the incoming wave by a directional coupler, amplified by $35$\,dB at $4$\,K and demodulated to baseband using homodyne detection. The complete circuit diagram of the experimental setup for qubit manipulation and dispersive readout is provided in Supplementary Note 1 and Supplementary Figure 1. 

\section{Data availability}
The data that support the findings of this study are available from the corresponding author upon reasonable request.

\section{Acknowledgements}
We thank K.D. Petersson, M.L.V. Tagliaferri, M.F. Gonzalez-Zalba and Y.-M. Niquet for fruitful discussions. The work was supported by the European Union’s Horizon 2020 research and innovation program under Grant Agreement No. 688539 MOS-QUITO (http://mos-quito.eu) and by the ERC Project No. 759388 LONGSPIN.

\section{Author contributions}
A.C. and R.E. performed the experiments with help from R.M.\,. A.C., R.M. and S.D.F. designed the experiment. R.L., L.H., B.B., M.V. fabricated the sample. A.C. analyzed the results with inputs from R.E, A. Apr\'a, A. Amisse, M.U., T.M., M.S., X.J, R.M. and S.D.F.\,. A.C., R.M. and S.D.F. wrote the manuscript. M.S., X.J, M.V. and S.D.F. initiated the project.

\section{Competing Interests}
The authors declare no competing interests.

\onecolumngrid
\vspace*{1in}

\setcounter{equation}{0}
\setcounter{figure}{0}
\section{SUPPLEMENTARY MATERIAL}

\renewcommand{\theequation}{S\arabic{equation}}  
\renewcommand{\thefigure}{S\arabic{figure}}

\section{Supplementary note 1: Measurement circuit}
\begin{figure}
\centering
	\includegraphics[width=\columnwidth]{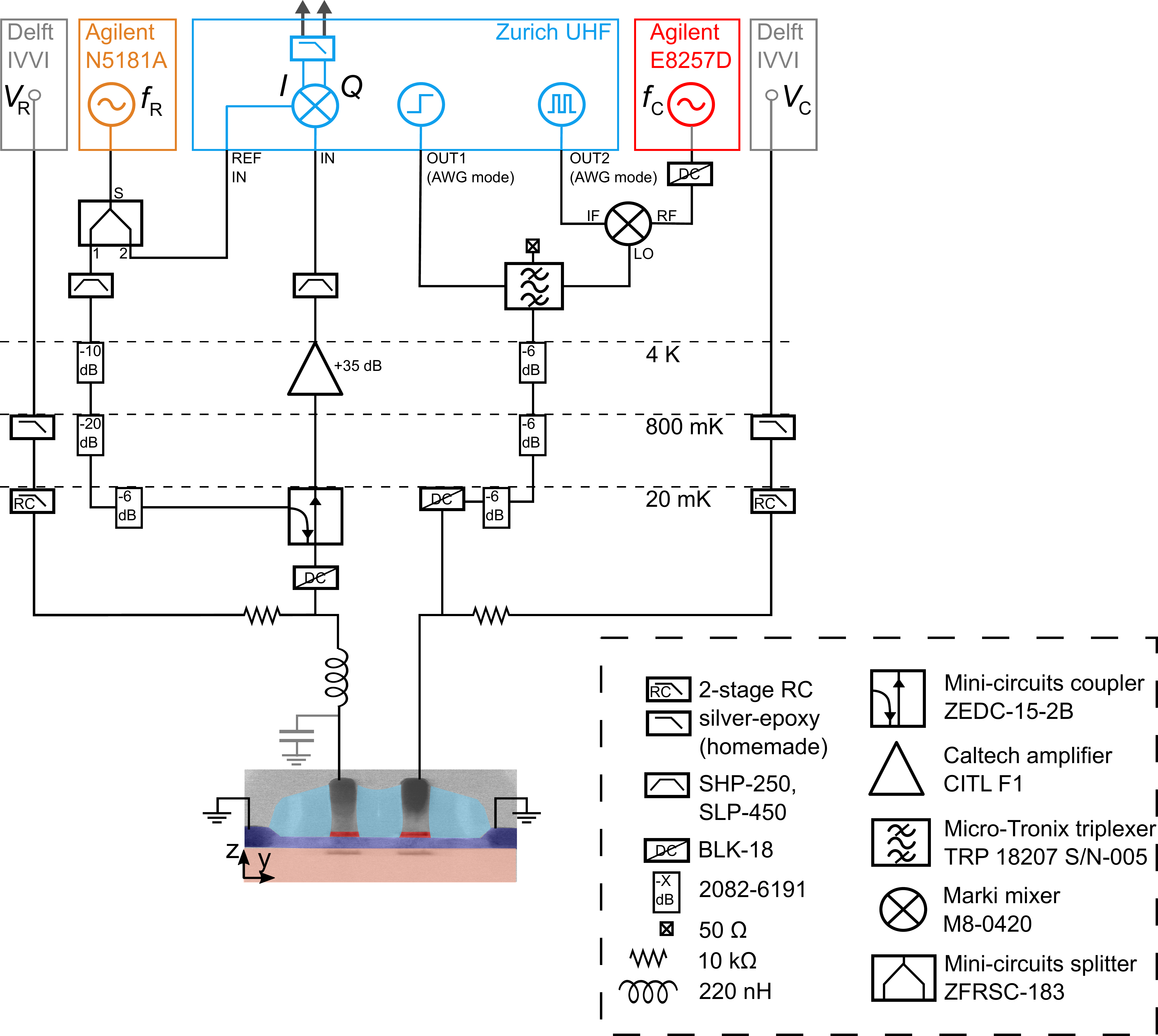} 
	\caption{Schematic of the qubit measurement setup. The circuitry of the right gate combines DC voltages for DQD electrostatic tuning, fast voltage pulses and EDSR microwave tones. For the left gate, the DC voltage is added to reflectometry radiofrequency signal for dispersive homodyne detection.}
	\label{fig:Fig1S}
\end{figure}

Supplementary Figure\,\ref{fig:Fig1S} shows the measurement circuitry of the experiment.\\
DC voltages are generated by room-temperature digital-to-analogue converters and filtered at low temperatures by home-made silver epoxy filters and 2-stage RC filters. These signals are applied to the two gate electrodes and source and drain contacts, the latter kept at 0\,mV throughout the whole experiment.\\
The reflectometry channel is fed by an Agilent N5181A RF source, which also provides the reference signal of a Zurich Ultra-High Frequency (UHF) lock-in for demodulation. The reflectometry tone is pass-band filtered at room temperature and attenuated at different stages of the fridge. It is added to the DC signal via a bias tee mounted on the sample holder. The tank circuit consists in a 220\,nH surface-mount inductor (Coilcraft 221XGLB) and a parasitic capacitance.
The reflected signal is separated from the incoming wave by a directional coupler and amplified at the 4\,K stage.\\
One output of the UHF is used in the AWG mode to precisely gate the microwave tone delivered by an Agilent E8257D source for coherent spin rotations. The resulting microwave bursts are added by a triplexer to the readout/manipulation pulses generated by the other UHF channel. The signal then passes through different attenuators and feeds a bias tee on the board.

\section{Supplementary note 2: Large stability diagram}
\begin{figure}
\centering
	\includegraphics[width=0.45\columnwidth]{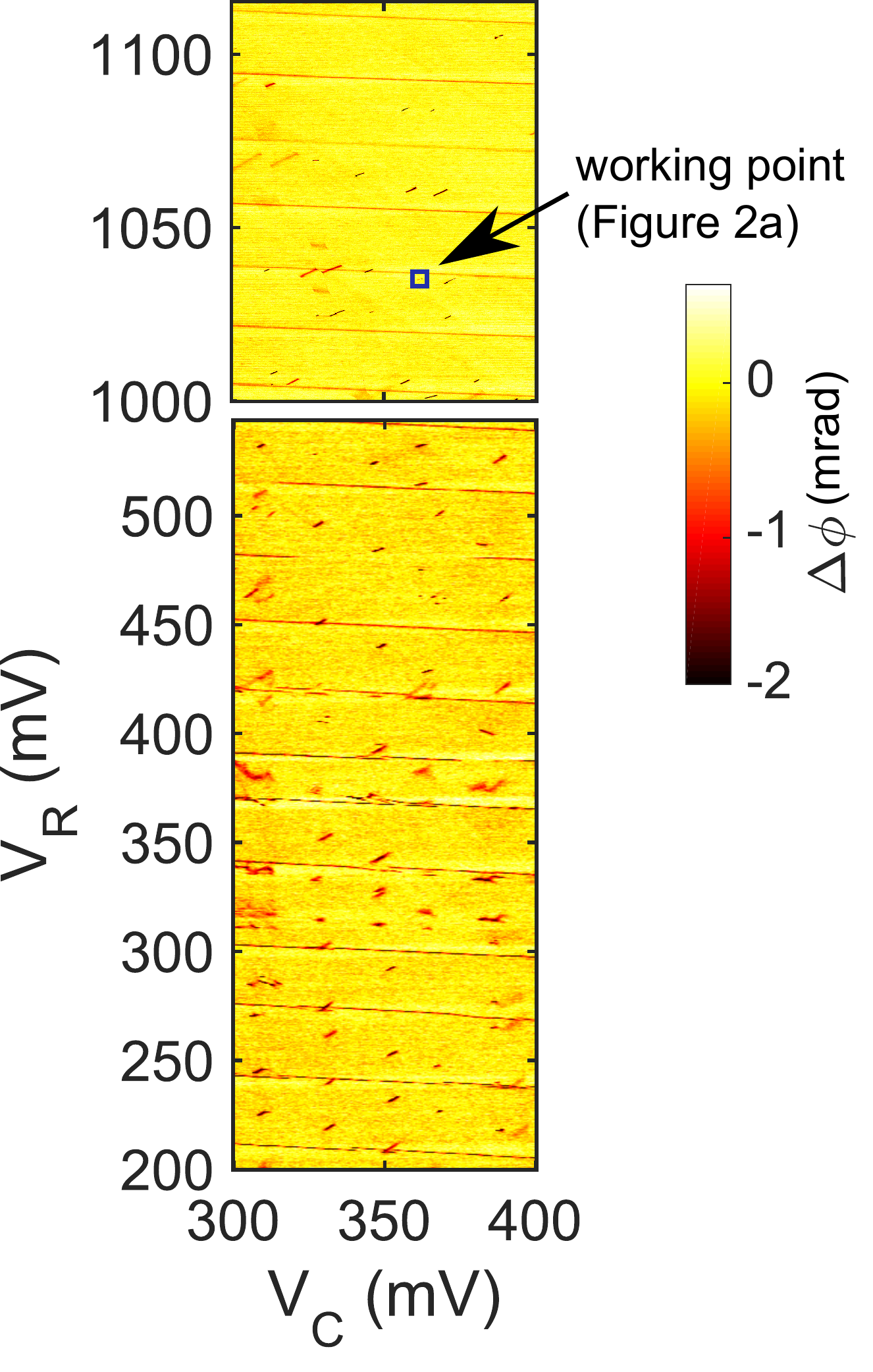} 
	\caption{Dispersively detected charge stability diagram of the device as a function of the two top gate voltages, $V_C$ and $V_R$. In the bottom panel, both gates are tuned in a strong accumulation mode, and the many hole regime, characterized by a regular arrangement of the interdot transition lines, is reached. In the upper panel, $V_R$ approaches the gate voltage threshold; as a result, interdot charge transitions are unequally placed. The blue square denotes the area zoomed-in in Fig.\,2a of the main text.}
	\label{fig:Fig2S}
\end{figure}
Supplementary Figure\,\ref{fig:Fig2S} shows two stability diagrams of the device under investigation. Both plots share the same $V_C$ voltage range. The other gate tunes the electrostatics of the channel from the many hole regime (bottom panel) to the voltage region we use to implement our qubit (top panel). In particular, the blue square highlights the area zoomed in the stability map of Fig.\,2a in the main text. 
Considering the diagram as a whole, two sets of features are present. First, a series of nearly horizontal parallel lines are visible. These lines repeat quite regularly from metallic DQDs to depletion, even the silicon channel is completely closed (data not shown). Consequently, we speculate that these features are related to the charging of objects extrinsic to the channel.\\
On top of this background, most of the short diagonal cuts on the yellow background are interdot transition lines.
The bottom part of Supplementary Fig.\,\ref{fig:Fig2S} reports the many hole regime where the voltage spacing between DQDs is approximately constant. The typical gate voltage between two charge states is about 25 mV. This value is consistent with other experiments on similar samples \cite{roro, gtensor}. 
Out of the many hole regime, the interdot lines are unevenly spaced, as displayed in the top panel. Importantly, for interdot tunnel couplings of few GHz (like the one studied in the main text), the interdot transition lines are quite thin in gate voltage, and are very likely not resolved in large maps obtained with large voltage steps. 
We use the threshold voltages at room temperature of the two gates and the addition voltage of the many hole regime for a rough estimation of the absolute filling of the dots. We obtain an order of magnitude of 5 holes and 10-20 holes in the left (i.e. mainly controlled by $V_R$) and right dot (mainly controlled by $V_C$), respectively.

\section{Supplementary note 3: Dispersive response of the DQD}
\subsection{Model}
In this Section we derive the spectrum the hole DQD close to the interdot charge transition presented in Fig.\,2a. As discussed in the main text, the dispersive signal is attributed to $("1,1") \leftrightarrow ("0,2")$ charge transfers. The superscripts " indicate that the numbers denote the parity-equivalent excess holes of the double dot.\\
%Holes in our dots have a mixed light and heavy character, as demonstrated by the detailed analysis of the g-factor anisotropy of Ref.\,[\citenum{gtensor}]. 
The excess charge of each dot is a qubit with spin-orbit eigenstates $\ket{\Uparrow}$ and $\ket{\Downarrow}$. 
The external magnetic field $B$ induces a Zeeman splitting between $\ket{\Downarrow}$ and $\ket{\Uparrow}$ equal to $g^*_{L(R)} \mu_B B$, with $g^*_{L(R)}$ the effective $g$-factor of the left (right) dot for a given direction of $B$ and $\mu_B$ the Bohr magneton.\\
We describe the DQD through the lowest five energy states in the basis $\{ \ket{\Uparrow \Uparrow}, \, \ket{\Uparrow \Downarrow}, \, \ket{\Downarrow \Uparrow}, \, \ket{\Downarrow \Downarrow}, \, \frac{1}{\sqrt{2}} \ket{0 \, (\Uparrow  \Downarrow -  \Downarrow \Uparrow)} \}$. The orbital spacing is $\sim 1$\,meV, which allows us to neglect the excited (triplet-like) $(0,2)$ state. The Hamiltonian then reads: 
\begin{equation}
    H=
    \begin{pmatrix}
    -\frac{1}{2}\epsilon + \frac{1}{2}(g^*_L+g^*_R) \mu_B B & 0 & 0 & 0 & 0 \\
    0 & -\frac{1}{2}\epsilon + \frac{1}{2}(g^*_L-g^*_R) \mu_B B & 0 & 0 & \frac{t}{\sqrt{2}} \\
    0 & 0 & -\frac{1}{2}\epsilon - \frac{1}{2}(g^*_L-g^*_R) \mu_B B & 0 & -\frac{t}{\sqrt{2}} \\
    0 & 0 & 0 & -\frac{1}{2}\epsilon - \frac{1}{2}(g^*_L+g^*_R) \mu_B B & 0 \\
    0 & \frac{t}{\sqrt{2}} & -\frac{t}{\sqrt{2}} & 0 & \frac{1}{2}\epsilon
    \end{pmatrix}
    .
\label{eq:H1}
\end{equation}
In this expression the DQD detuning $\epsilon$ is with respect to the middle point energy between the eigenstates at $B=0$\,T. The tunnel coupling $t$ connects the antiparallel spin states $\ket{\Downarrow \Uparrow}$ and $\ket{\Uparrow \Downarrow}$ to the singlet $S(0,2) = \frac{1}{\sqrt{2}} \ket{0 \, (\Uparrow  \Downarrow -  \Downarrow \Uparrow)}$, thereby allowing interdot charge transitions.\\
The DQD spectrum in Fig.\,2d of the main text displays the eigenvalues of Supplementary Eq.\,\ref{eq:H1} as a function of $\epsilon$ with $g^*_L=1.62$, $g^*_R=2.12$, $t=8 \, \mu$eV and $B=0.65$\,T.

\subsection{DQD dispersive response}

To clarify the origin of the dispersive signal close to the interdot transition line, we map the DQD of Supplementary Eq.\,\ref{eq:H1} onto a singlet-triplet basis $\{ T_+(1,1), \, T_0(1,1), \, T_-(1,1), \, S(1,1), \, S(0,2)\}$:
\begin{equation}
    H'=
    \begin{pmatrix}
    -\frac{1}{2}\epsilon + \frac{1}{2}(g^*_L+g^*_R) \mu_B B & 0 & 0 & 0 & 0 \\
    0 & -\frac{1}{2}\epsilon & 0 & \frac{1}{2}(g^*_L-g^*_R)\mu_B B & 0 \\
    0 & 0 & -\frac{1}{2}\epsilon - \frac{1}{2}(g^*_L+g^*_R) \mu_B B & 0 & 0 \\
    0 & \frac{1}{2}(g^*_L-g^*_R)\mu_B B & 0 & -\frac{1}{2}\epsilon & t\\
    0 & 0 & 0 & t & \frac{1}{2}\epsilon
    \end{pmatrix}
    .
\label{eq:H2}
\end{equation}
The singlet states have a curvature due to coupling term $t$, which leads to a non-zero quantum capacitance and a consequent dispersive shift of the resonant frequency. Concerning $T_0(1,1)$, it is usually not dispersively sensed since its second derivative with respect to $\epsilon$ is zero. However, Fig.\,2e shows a finite phase response for $T_0(1,1)$. It comes from the electric dipole due to the coupling with $S(1,1)$ via $\frac{1}{2}(g^*_L-g^*_R) \mu_B B$, which eventually implies a second-order coupling with $S(0,2)$.\\
\\
We model the DQD dispersive response by quantum capacitance contributions with a Boltzmann distribution \cite{Mizuta, Petta_parity}.
%Effects of strong rf-driving by the probing frequency $f_R$ can be neglected. The smallest avoided crossing of Eq.\,\ref{eq:H2} is among the hybridized singlets $S_g$ and $T_0$, which is of the order of $t$ ($\sim 2$\,GHz), therefore higher than $f_R=339$\,MHz.\\
Fast excitations/relaxations in the singlet manifold may contribute to the DQD phase response through the tunnel capacitance \cite{Mizuta}. However, if such nonadiabatic processes are slow ($\sim 100$\,MHz) compared to the probing frequency $f_R$, the tunnel capacitance is small with respect to the quantum capacitance; on the other hand, with fast charge relaxations ($\sim 1$\,GHz) the interdot ridge would have lineshape and width not compatible with the magnetic field evolution reported in Fig.\,2.\\ 
\\
In the basis set of Supplementary Eq.\,\ref{eq:H2}, the spin-orbit (SO) transition matrix elements are supposed weak compared to $t$ and the Zeeman terms. Sizable spin-flip tunnelling terms like $t_{SO}^{\ket{T_-}} \ket{T_-(1,1)} \bra{S(0,2)}$ and $t_{SO}^{\ket{T_+}} \ket{T_+(1,1)} \bra{S(0,2)}$ would lead to a dispersive signal with a strong magnetic field dependence. We found no evidence of the corresponding dispersive signals in the magnetospectroscopy data discussed in the main text.\\
A coupling factor $t_{SO}^{\ket{T_0(1,1)}}$ between $T_0(1,1)$ and $S(0,2)$ comparable to $t$ has not to be expected neither. From simulations at $B>0.5$\,T, with $t_{SO}^{\ket{T_0}} \sim t$ the phase resonance of the interdot transitions would resemble a pronounced peak with a bearly-visible shoulder on the right edge, not consistent with data in Figs.\,2 and 3 of the main text.
However, we can’t rule out these such spin-flip tunneling terms might be relevant for orientations of the external magnetic field different from those investigated here.\\
\\
A comprehensive description of the experimental phase signal is achieved by considering the excited levels of the DQD as partially populated.
Each state leads to an averaged phase signal $\langle \Delta \phi \rangle_i = \Delta \phi_i^{T=0} \cdot e^{-E_i \beta} /Z$, where $\Delta \phi_i^{T=0}$ is proportional to the quantum capacitance of the state with energy $E_i$ in the 0\,K limit, $\beta=1/k_B T_{\text{eff}}$ with $k_B$ Boltzmann constant and $T_{\text{eff}}$ an effective temperature, and $Z$ is the partition function over the five states.
The measured phase signal then is $\Delta \phi = \sum_i \langle \Delta \phi \rangle _i$.\\
The coupling term $t$ is estimated from a detuning trace at $B=0$\,T. According to the model just described, the full width at half maximum (FWHM) of the phase interdot ridge as a function of the effective temperature $T_{\text{eff}}$ evolves as shown in Supplementary Figure\,\ref{fig:Fig3S}.

\begin{figure}
\centering
	\includegraphics[width=0.45\columnwidth]{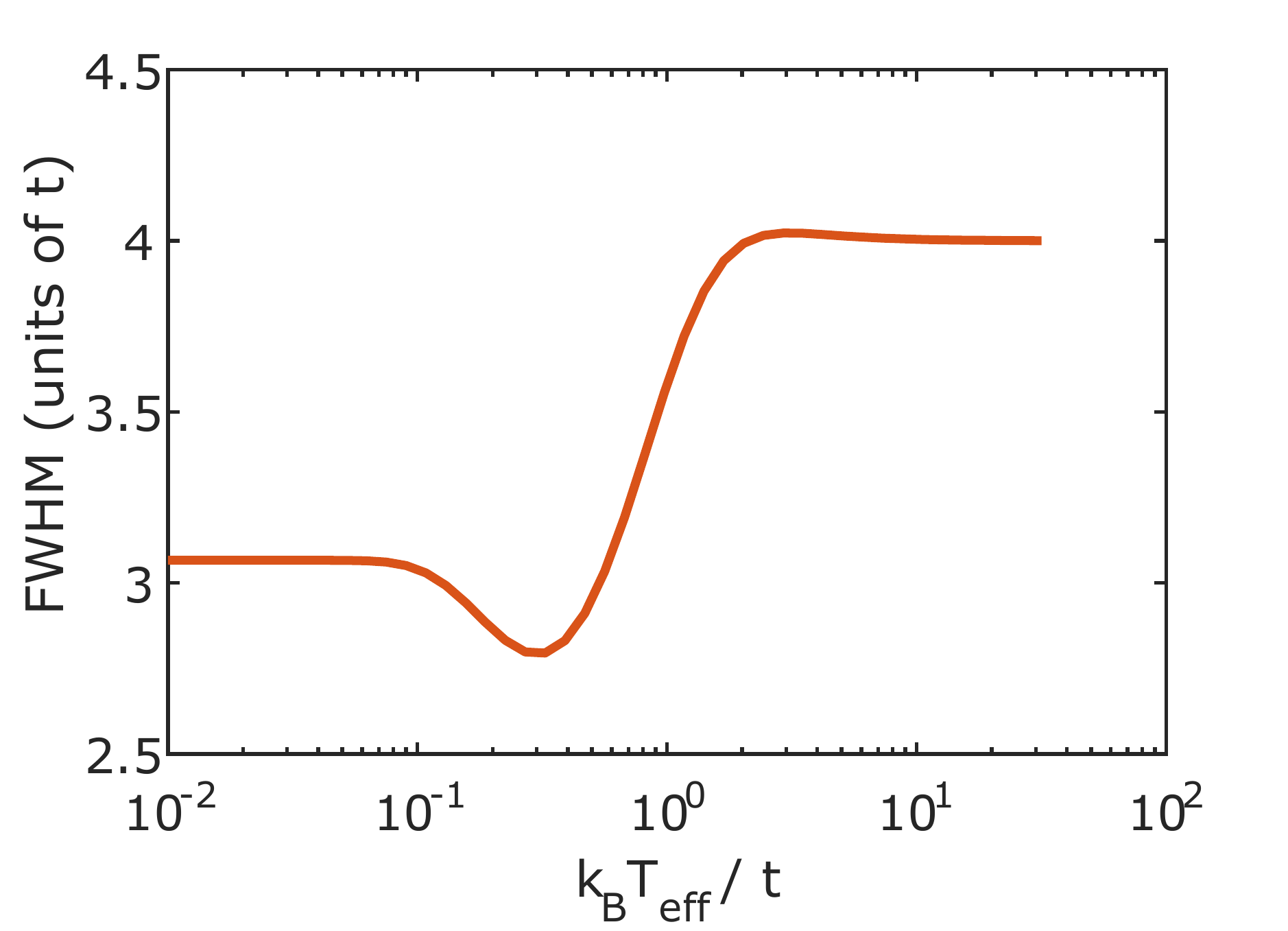} 
	\caption{Predicted evolution of the full width at half maximum (FWHM) of the interdot dispersive resonance as a function of temperature at $B=0$\,T. The phase signal is supposed composed solely of quantum capacitance contributions.}
	\label{fig:Fig3S}
\end{figure}

Two limiting situations are envisaged. At low temperature, $k_B T_{\text{eff}} < t/10$ and the width of the interdot signal is set by the tunnel coupling to $\sim 3t$. Here just the ground singlet is populated. In the opposite limit of high temperature, $k_B T_{\text{eff}} > 2t$, the threefold  triplet and both bonding and anti-bonding singlet are thermally populated; by sweeping $T_{\text{eff}}$, the magnitude of the interdot resonance drops, but the FWHM saturates at $\sim 4t$. In the intermediate regime, the FWHM increases progressively with $T_{\text{eff}}$, up to the saturation point occurring at $k_B T_{\text{eff}} \simeq 2t$.\\
Furthermore, Supplementary Fig.\,\ref{fig:Fig3S} demonstrates that the FWHM allows to estimate $t$ in the $(3t,\,4t)$ range whatever the temperature is. This distinguishes dispersive readout from charge sensing (especially when $k_B T_{\text{eff}} > 2t$), as the resonator sensitivity is ultimately constrained to the avoided crossings in the energy level diagram.\\
Fits to the interdot detuning phase shift yield $t=8.5$\,$\mu$eV and $t=6.4$\,$\mu$eV in the low and high temperature limit, respectively. The evolution of the interdot transition line versus the magnetic field is reproduced qualitatively assuming the lowest tunnel coupling and 0.25\,K as effective temperature.
The inset in Fig.\,2b of the main text is obtained with $g^*_L=1.52$, $g^*_R=2.02$, $t=6 \, \mu$eV and $T_{\text{eff}}=0.25$\,K. The one-dimensional cuts in Fig.\,2f are taken at $B=0, \,0.35, \,0.5$ and 0.85\,T.\\

\section{Supplementary note 4: Detuning position for dispersive readout}

\begin{figure}
\centering
	\includegraphics[width=0.5\columnwidth]{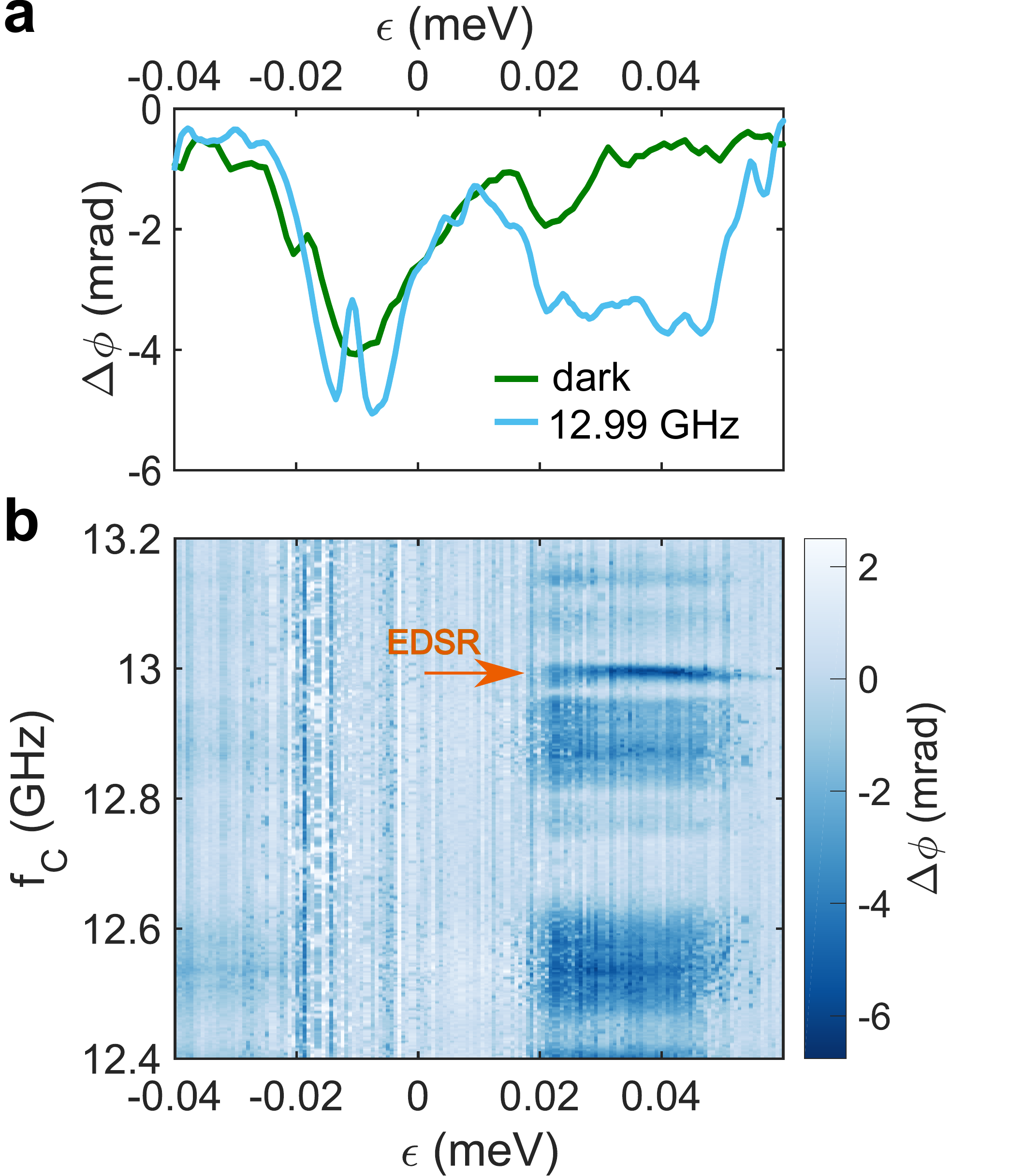} 
	\caption{(a) One-dimensional detuning scan of the interdot line without microwave radiation ('dark') and with $f_C=12.99$\,GHz applied. The right peak (associated to $\ket{T_0}$) is enhanced and largely broadened when EDSR transitions take place. (b) Colorplot of the phase response as a function of the detuning and the driving frequency. The map is acquired by sweeping $f_C$ and stepping $\epsilon$; at the beginning of each line, the phase of the reflectometry signal is set to 0. Signal related to EDSR is indicated by the orange arrow.}
	\label{fig:Fig4S}
\end{figure}

Inset of Fig.\,3c shows the behavior of the phase signal we expect when second harmonic EDSR transitions are promoted between the $\ket{T_-}$ and $\ket{T_0}$ states at point $I/R$ (we write the states as kets from now on). From our model, we set $g^*_L=1.575$, $g^*_R=2.075$, $t=6\,\mu$eV, $B=0.52$\,T and $T_{\text{eff}}=0.25$\,K. We also impose $f_C=7.42$\,GHz, as in the experimental trace of the main panel. 
We find that the resonant condition $f_C = |E_{\ket{T_0}} - E_{\ket{T_-}}|/2h$ is met at finite $\epsilon$ ($23.5\,\mu$eV), in agreement with data.
At this detuning, we model the EDSR peak by equalizing the occupation probability of both states to $[P_{\ket{T_-}}+P_{\ket{T_0}}]/2$, where $P_{\ket{T_-}}=\exp{(-E_{\ket{T_-}} \beta})/Z$ and $P_{\ket{T_0}}=\exp{(-E_{\ket{T_0}} \beta})/Z$.
The phase response is then convolved with a Gaussian distribution with a variance of $50 \, (\mu \text{eV})^2$ accounting for the observed detuning broadening.
One might expect the detuning position of the EDSR peak to depend on $f_C$, along with an increase of phase signal with approaching to $\epsilon=0$. However, as observed in other types of Si qubits \cite{Jiang_ValleyQubit, Hybrid_extending, PettaLZS}, in the vicinity of $\epsilon=0$ decoherence rates increase as well, which limits the detuning window for convenient reflectometry readout.\\
Supplementary Figure\,\ref{fig:Fig4S} shows the phase response as a function of detuning (along $V_C$) and EDSR driving frequency. The $B$ field is oriented as in the qubit measurements of Fig.\,4 of the main text.  
The phase signal due to EDSR transitions is highlighted by the orange arrow of panel b), while the other nonzero phase features are due to spurious photon-assisted events or noise.\\
As pointed out in the main text, the microwave-induced population of $\ket{T_0}$ state is detected in a "shallow" $(1,1)$ charge stability region ($\epsilon > 0.02$\,meV) with a nearly constant dispersion $d|E_{\ket{T_0}} - E_{\ket{T_-}}|/d\epsilon$. In this regime, the qubit is robust with respect to fluctuations in the energy splitting, as testified by the relaxation time in the microsecond range.
Closer to the alignment of the electrochemical potentials of the two dots ($0 < \epsilon < 0.02$\,meV), the EDSR signal is not resolved, probably due to inhomogeneous energy broadening between the resonant states. This makes this bias regime unsuitable for readout.  
% maybe also a strong increase of the relaxation rate.

\section{Supplementary note 5: $\boldsymbol{T_1}$ estimation}

To extract the spin relaxation time at the readout position $I$ of Fig.\,4b, we use a pulse length of 250\,ns and sweep $t_{\text{wait}}$, see Fig.\,4a. During the pulse, a microwave burst of 100\,ns flips one of the two spins. We normalize the amplitude of the phase shift by a factor $(1+250\,\text{ns}/t_{\text{wait}})$ since the signal is acquired during the whole period $T_M$.\\
The readout projects $\ket{\Uparrow \Downarrow}$ on the $\{\ket{T_0},\,\ket{T_-} \}$ basis. The time-dependent probability that the spin relaxes in $\ket{T_-}$ is given by $P(t)_{\ket{T_-}}=P(t=0)_{\ket{T_0}}\exp{(-t/T_1)}$. The time averaged data points in Fig.\,4b are then fitted to $\Delta \phi = a_0 - a_1 \frac{T_1}{T_M} \big( \exp{(-T_M/T_1)} -1 \big) $, with $T_M=t_{\text{wait}}+\text{250\,ns}$.

\twocolumngrid

\bibliography{biblioReflectoQubitArxiv}

\end{document}